\documentclass{PoS}

\usepackage{amsmath}
\usepackage{graphicx}
\usepackage{mathrsfs}

\newcommand{\de}{\delta} 
\newcommand{\eref}[1]{Eq.~(\ref{#1})}
\newcommand{\fref}[1]{Fig.~\ref{#1}}

\title{Going beyond the propagators of Landau gauge Yang-Mills theory}

\ShortTitle{Going beyond the propagators of Landau gauge Yang-Mills theory}

\author{\speaker{Markus Q. Huber}, Lorenz von Smekal\\
	Institut f\"ur Kernphysik,
	Technische Universit\"at Darmstadt,
	Schlossgartenstr. 2,
	64289 Darmstadt, Germany\\
	E-mail: \email{markus.huber@physik.tu-darmstadt.de},
\email{lorenz.smekal@physik.tu-darmstadt.de}
}

\abstract{
We present results for the propagators and the ghost-gluon vertex of Landau gauge Yang-Mills theory obtained from Dyson-Schwinger equations. Solving these three quantities simultaneously constitutes a new step in truncating these equations. We also introduce a new model for the three-gluon vertex that is motivated by lattice results. It features a zero crossing which is confirmed a posteriori by a Dyson-Schwinger calculation. Within our setup we can reproduce lattice data very well. We establish that also for the ghost-gluon vertex a difference between decoupling and scaling solutions is present. For the scaling solution we discuss the possibility of modifying the infrared exponents via an angle dependence of the ghost-gluon vertex. However, no such dependence is found in our calculations. Finally, we calculate the Schwinger function for the gluon propagator.
}

\FullConference{Xth Quark Confinement and the Hadron Spectrum,\\
		October 8-12, 2012\\
		TUM Campus Garching, Munich, Germany}

\begin{document}

\section{Two-point functions}

Extracting non-perturbative information from a quantum field theory is in most cases a hard task. One possibility is to calculate its Green functions which are related by several sets of functional equations. A concrete example of such a set are the equations of motion of the Green functions, called Dyson-Schwinger equations (DSEs) \cite{Alkofer:2000wg,Alkofer:2008nt}. An inherent problem is that there are infinitely many equations. The usual way to make quantitative calculations is to neglect some Green functions and/or model others. The remaining ones are then solved self-consistently. The natural way to test the reliability of such a truncated system is to compare it with a larger truncation. However, in practice this is a difficult task and it can take some time until such a calculations becomes feasible both from the conceptual and from the computational point of view. An example for this is provided below. Nevertheless, it has to be stressed that raising the number of included Green functions can allow a systematic improvement. In practice, however, it is more common to estimate truncation artifacts from comparisons with other methods.

Sometimes we are in an even better situation and can estimate the influence of truncations. The most prominent example is surely the high energy regime of QCD. Since the strong interaction is asymptotically free, a perturbative loop expansion provides a reliable approximation which can be improved systematically by including higher powers of the coupling. On the other end of the energy spectrum, in the infrared (IR), it can also be possible to determine a clear hierarchy of leading and suppressed contributions, if a so-called scaling solution is realized \cite{Alkofer:2004it,Fischer:2009tn,Huber:2010ne}. Then the qualitative behavior of all Green functions can be determined in form of IR power laws. Landau gauge Yang-Mills theory may actually possess such a solution \cite{vonSmekal:1997is,vonSmekal:1997vx,Fischer:2008uz,Huber:2012kd}. Knowledge of additional information like this is of course a useful guideline in constructing truncations.

In the present work we consider Landau gauge Yang-Mills theory. It is illustrative in the light of the discussion above to recall the development of truncations in this case. In a very basic approximation the ghosts and all four-point functions were neglected \cite{Mandelstam:1979xd}. The only DSE to be solved is the gluon propagator DSE. The resulting gluon propagator is IR divergent, which is nowadays known to be wrong. The next step is the inclusion of ghosts, so that a coupled system of two DSEs has to be solved. This step was taken 18 years later \cite{vonSmekal:1997is,vonSmekal:1997vx} and resulted in a drastic change, as the gluon dressing became IR suppressed. Based on analytic and numeric results from various methods, we can be confident that a truncation at this level yields qualitatively reliable results.

Within this class of truncations two types of solutions can be obtained. However, this is not related to the truncation itself but to the boundary conditions imposed on the DSEs and a difference is only visible at small momenta  \cite{Fischer:2008uz}. The decoupling type consists of a family of solutions which have an IR finite ghost dressing and an IR finite gluon propagator \cite{Fischer:2008uz,Dudal:2008sp,Boucaud:2008ji,Aguilar:2008xm,Alkofer:2008jy}. The scaling solution, on the other hand, features an IR divergent ghost dressing and an IR vanishing gluon propagator \cite{vonSmekal:1997is,vonSmekal:1997vx,Fischer:2008uz,Huber:2012kd}. It is still under debate if both solutions exist, especially since all "standard" lattice calculations recover the decoupling type of solutions. However, lattice results do depend on the method used for sampling Gribov copies, see, e.g., \cite{Sternbeck:2006rd,Maas:2011se}. Recently a way was found to sample Gribov copies such that the results are modified as expected from functional equations \cite{Sternbeck:2012mf}.

In the following we will present results from a new truncation that goes beyond the propagator level by including the ghost gluon-vertex dynamically and employs a model for the three-gluon vertex in agreement with lattice results \cite{Huber:2012kd}.

\section{Three-gluon vertex}

\begin{figure}[tb]
 \begin{center}
  \includegraphics[width=0.49\textwidth]{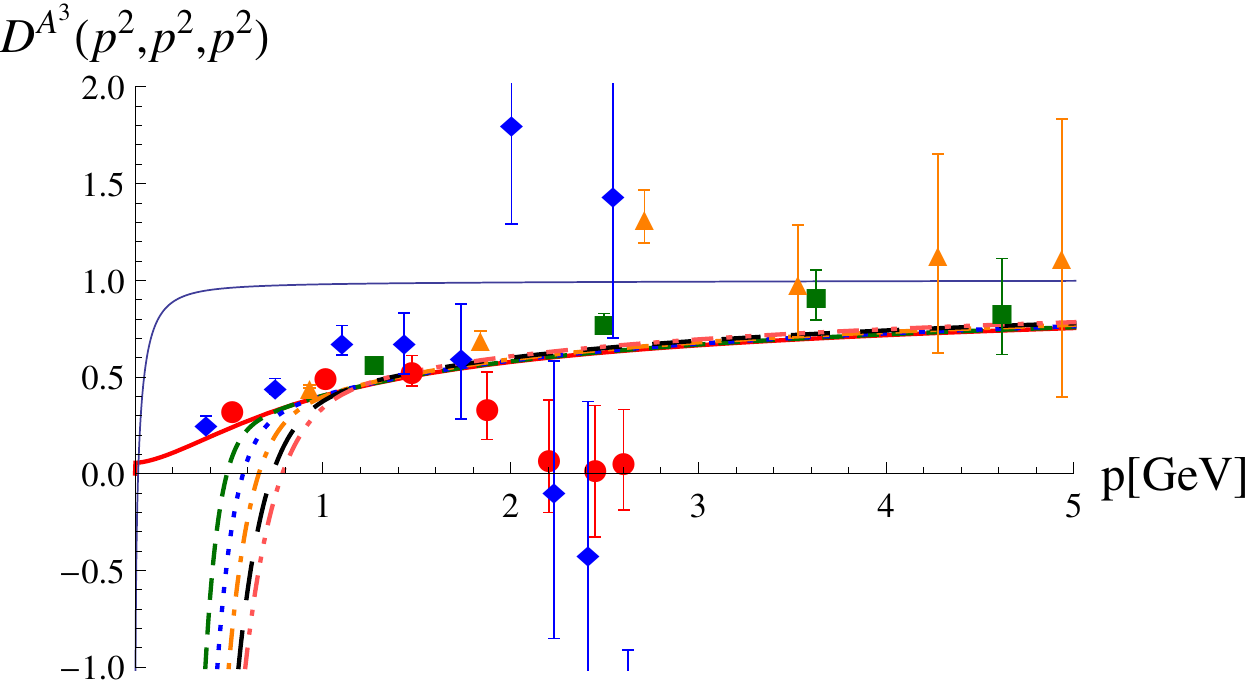}
  \includegraphics[width=0.49\textwidth]{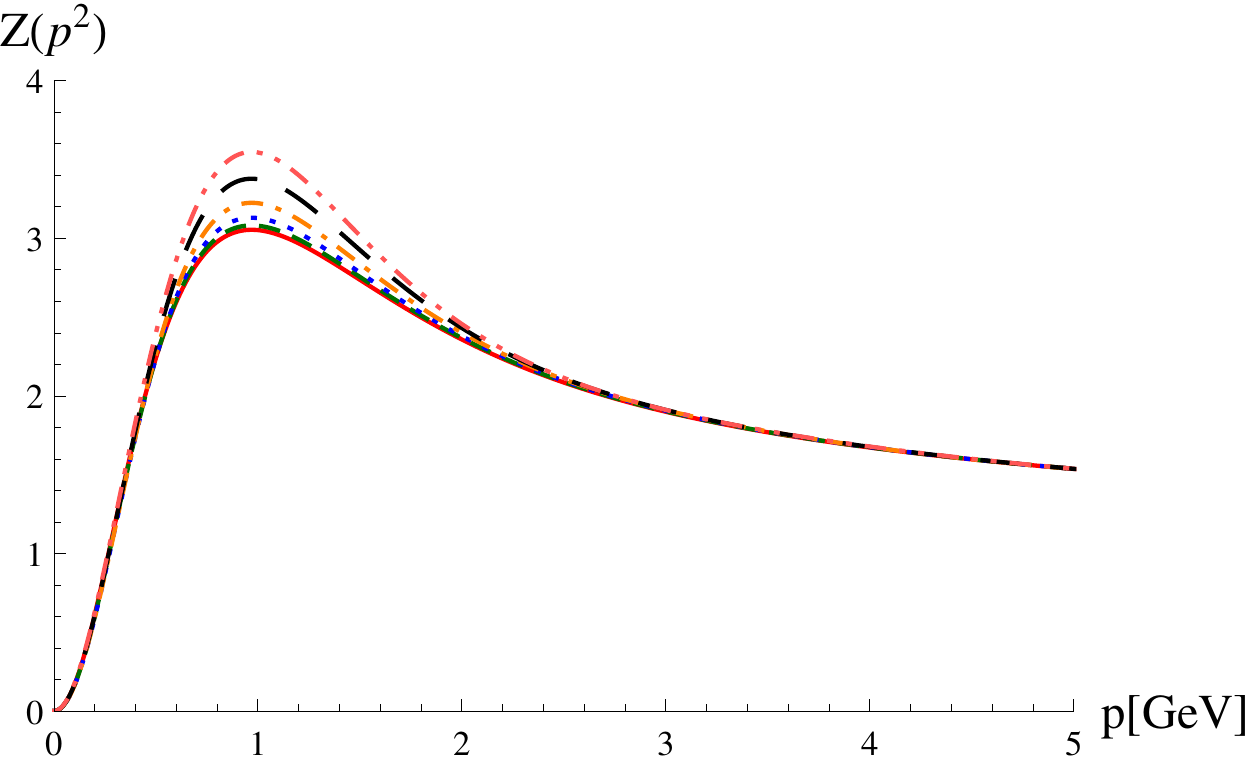}
  \caption{\label{fig:comp3gModel}Comparison of different parameters for the three-gluon vertex model: $h_{IR}=-1$, $\Lambda_{3g}=0,\,0.86,\,1.08,\,1.31,\,1.53,\,1.7\,GeV$.
\textit{Left:} Symmetric configuration of the vertex model. Thick lines from left to right correspond to rising values of the parameter $\Lambda_{3g}$. The thin line is the IR leading contribution of the three-gluon vertex calculated from its DSE showing a zero crossing at very low momenta. At zero momentum it is finite. Individual points are lattice data from ref.~\cite{Cucchieri:2008qm}: Red circles correspond to $N=16$/$\beta=2.2$, green  squares to $N=16$/$\beta=2.5$, blue diamonds to $N=22$/$\beta=2.2$ and orange triangles to $N=22$/$\beta=2.5$.
\textit{Right:} The corresponding gluon dressing function. Curves from bottom to top correspond to values of $\Lambda_{3g}$ from low to high.}
 \end{center}
\end{figure}

For the three-gluon vertex we employ the following model motivated by lattice calculations:
\begin{align}\label{eq:3g}
 D^{A^3}(p,q,-p-q)=G\left(\frac{x+y+z}{2}\right)^{\alpha}Z\left(\frac{x+y+z}{2}\right)^{\beta}+h_{IR} \,G(x+y+z)^{3}(f^{3g}(x)f^{3g}(y)f^{3g}(z))^4,
\end{align}
with the damping factors
\begin{align}
 f^{3g}(x):=\frac{\Lambda^2_{3g}}{\Lambda_{3g}^2+x}.
\end{align}
We choose $h_{IR}<0$ so that the vertex becomes negative for low momenta. This is observed with Monte Carlo simulations in two and three dimensions \cite{Maas:2007uv,Cucchieri:2008qm}, but not in four \cite{Cucchieri:2008qm}, probably because the lattice data does not reach far enough into the IR. We performed a DSE calculation for the vertex based on the results obtained for the propagators and the ghost-gluon vertex and taking into account only the IR leading diagram. We indeed found a zero crossing, see \fref{fig:comp3gModel}, which is compatible with the available lattice data. The position of the crossing may still change in a more advanced truncation, but it is very unlikely that it will vanish. This is corroborated by results in two dimensions, where more detailed calculations were performed and the influence of other diagrams was found to be negligible in the IR \cite{Huber:2012zj}.

In \eref{eq:3g} the IR leading part is damped out for larger momenta by the damping functions $f(x)$. The other term is motivated by perturbation theory: By choosing $\alpha$ and $\beta$ appropriately, we reproduce the anomalous dimension of the three-gluon vertex, which is $\gamma_{3g}=1+3\delta=17/44$ with $\delta=-9/44$ the ghost anomalous dimension. The second condition to fix $\alpha$ and $\beta$ is IR finiteness. We obtain for the scaling solution $\alpha=-2-6\de$ and $\beta=-1-3\de$ and for decoupling $\alpha=3+1/\de$ and $\beta=0$. In order to produce the correct anomalous dimensions for the propagators, it is necessary to make an adjustment to the gluon loop in the gluon DSE, where the renormalization constant $Z_1$ of the three-gluon vertex appears. We replace it by a momentum-dependent renormalization group improvement term that equals the UV part of the three-gluon vertex model \cite{vonSmekal:1997vx}. Such a modification is unavoidable in order to obtain the correct anomalous dimensions, because they result from the resummation of infinitely many diagrams which cannot be taken into account directly in the present approach.

In \fref{fig:comp3gModel} results for the gluon dressing function for varying values of $\Lambda_{3g}$ are shown. The influence of the zero crossing is clearly visible: If its position moves to higher momenta, the bump in the gluon dressing function rises. In general the model shows good agreement with the lattice data, which is, however, for two colors while our calculation is for three colors. From \fref{fig:comp3gModel} we conclude that the zero crossing, if chosen at a value in agreement with lattice data, has no influence on the propagators. The reason is that in the gluon loop the vertex is multiplied with two gluon propagators which are IR suppressed. Besides the IR part also the Bose symmetry of the vertex has an effect on the mid-momentum regime of the propagators, see ref.~\cite{Huber:2012kd}. In the following we will use the three-gluon vertex model to mimic effects of the two-loop diagrams in the gluon DSE by choosing the zero crossing such that the propagators agree as good as possible with lattice data. The model with the corresponding choice of parameters is called optimized effective vertex.

\section{Ghost-gluon vertex}

The ghost-gluon vertex is an important quantity in modern truncation schemes of functional equations, because it is the only primitively divergent vertex via which the ghost interacts. The reason for the success of modern truncation schemes is that this vertex has only a very mild momentum dependence. Thus approximating it by its bare expression produces reasonable results. Non-perturbative modifications of the vertex were investigated with various approaches: Monte-Carlo simulations \cite{Sternbeck:2006rd,Maas:2007uv,Cucchieri:2008qm}, semi-perturbative DSE calculations \cite{Schleifenbaum:2004id,Alkofer:2008dt}, OPE analysis \cite{Boucaud:2011eh} and based on this indirect studies via effects in the ghost DSE \cite{Dudal:2012zx}. Here we will include the vertex dynamically, i.e., we solve the DSEs of the two propagators and the vertex simultaneously. We employ the following basis for the vertex:
\begin{align}
 \Gamma^{A\bar{c}c,abc}_\mu(k;p,q):=f^{abc}\Gamma^{A\bar{c}c}_\mu(k;p,q):=i\,g\,f^{abc}\left(p_\mu A(k;p,q)+k_\mu B(k;p,q)\right),
\end{align}
where the momenta $k$, $p$ and $q$ refer to the gluon, the anti-ghost and the ghost, respectively. Due to the transverse gluon propagator of Landau gauge, only the dressing $A(k;p,q)$ is relevant. The corresponding DSE as well as those for the propagators can be found in ref.~\cite{Huber:2012kd}.

\begin{figure}[tb]
 \begin{center}
  \includegraphics[width=0.49\textwidth]{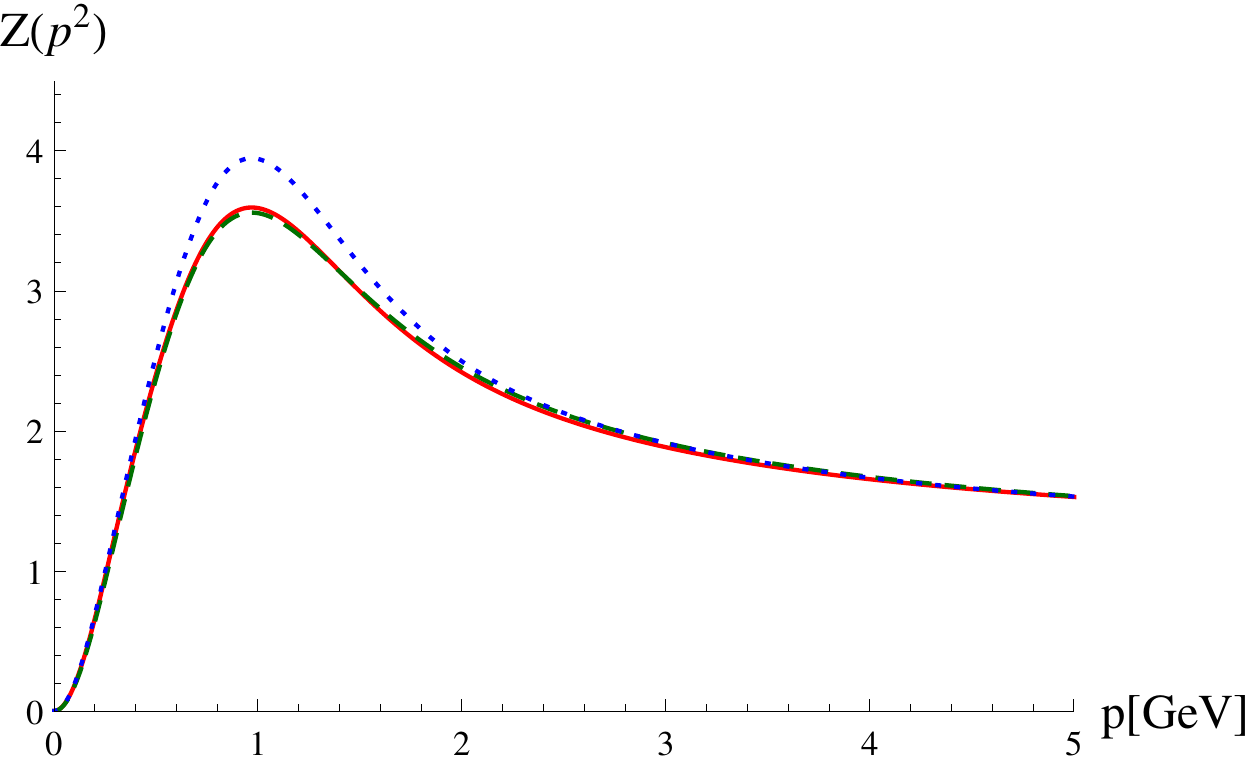}
  \includegraphics[width=0.49\textwidth]{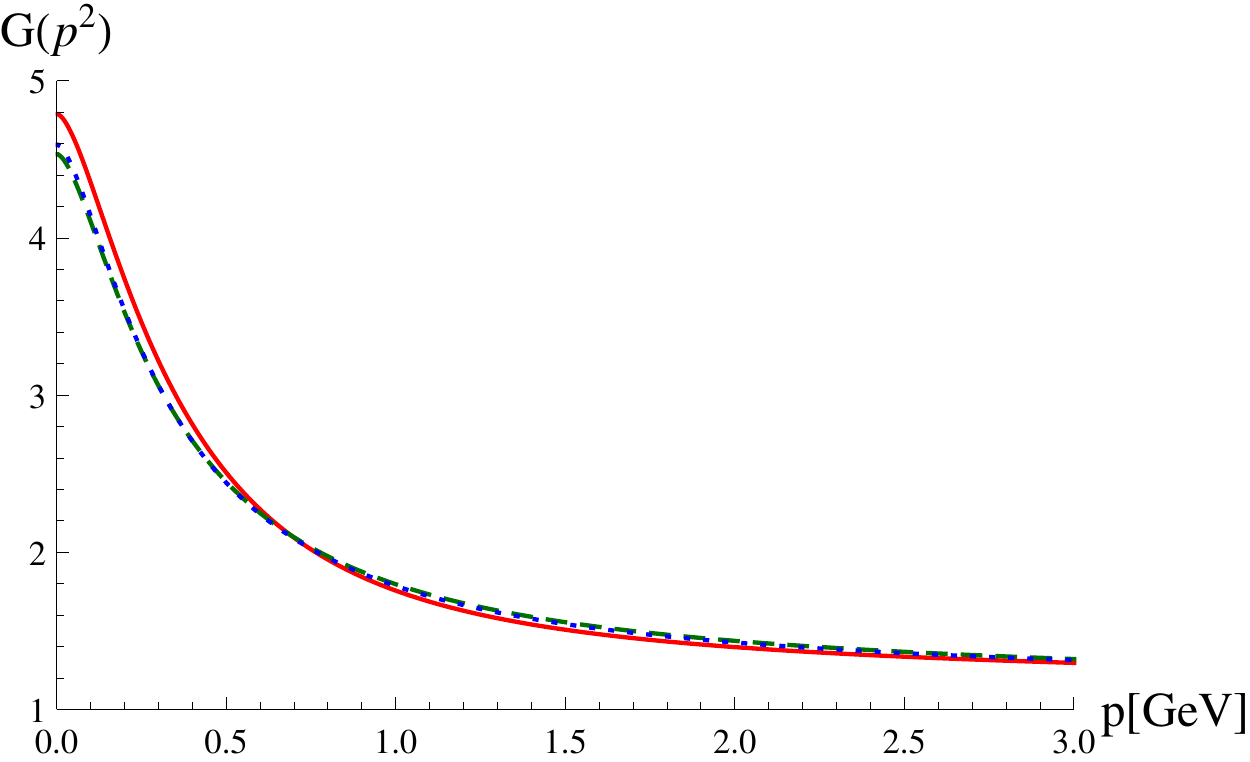}
  \caption{\label{fig:compOptEff3g_props}Comparison of the gluon propagator dressing function $Z(p^2)$ and the ghost dressing function $G(p^2)$ for different setups. Red/continuous line: dynamic ghost-gluon vertex, $\Lambda_{3g}=2.1\,GeV$. Green/dashed line: bare ghost-gluon vertex, $\Lambda_{3g}=1.8\,GeV$. Blue/dotted line: bare ghost-gluon vertex, $\Lambda_{3g}=2.1\,GeV$.}
 \end{center}
\end{figure}

The effect of the dynamic ghost-gluon vertex is illustrated in \fref{fig:compOptEff3g_props}, where calculations with a bare ghost-gluon vertex are compared to one with a dynamic vertex. The red/continuous line corresponds to the solution with the optimized effective three-gluon vertex and a dynamic ghost-gluon vertex. The corresponding calculation with the same parameters for the three-gluon vertex but a bare ghost-gluon vertex is represented by the blue/dotted line. However, although there is quite some change in the mid-momentum regime of the gluon propagator, this only provides a qualitative picture, since the model for the three-gluon vertex depends on the propagators and is thus different in the two calculations. Finally, we note that it is possible to approximately accommodate the modifications introduced by a dynamic vertex also in the three-gluon vertex model. This can be seen from the green/dashed line, for which the parameters of the model were adjusted such that the gluon propagator from the lattice is reproduced using a bare ghost-gluon vertex.

\section{Results}

Technical details on the calculations can be found in ref.~\cite{Huber:2012kd}. The programs \textit{DoFun} \cite{Alkofer:2008nt,Huber:2011qr} and \textit{CrasyDSE} \cite{Huber:2011xc} were used for deriving and solving the DSEs.

\begin{figure}
 \begin{center}
  \includegraphics[width=0.49\textwidth]{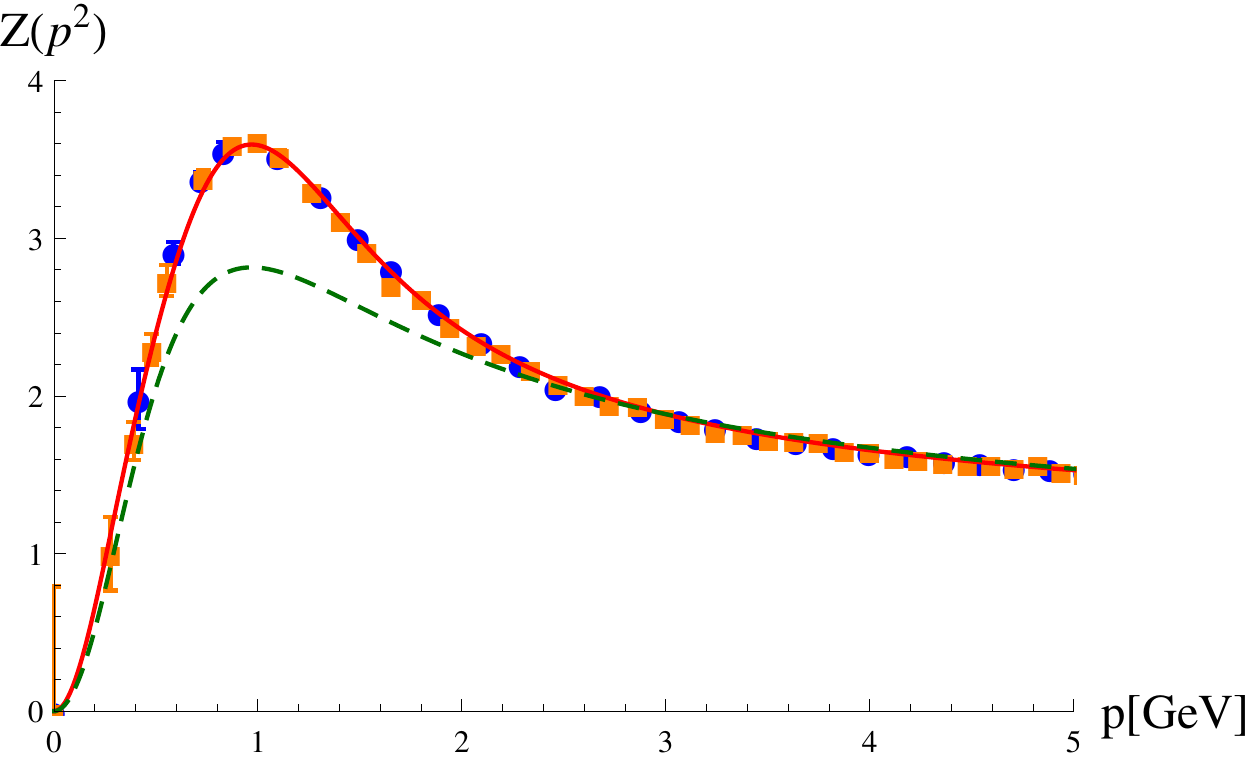}
  \includegraphics[width=0.49\textwidth]{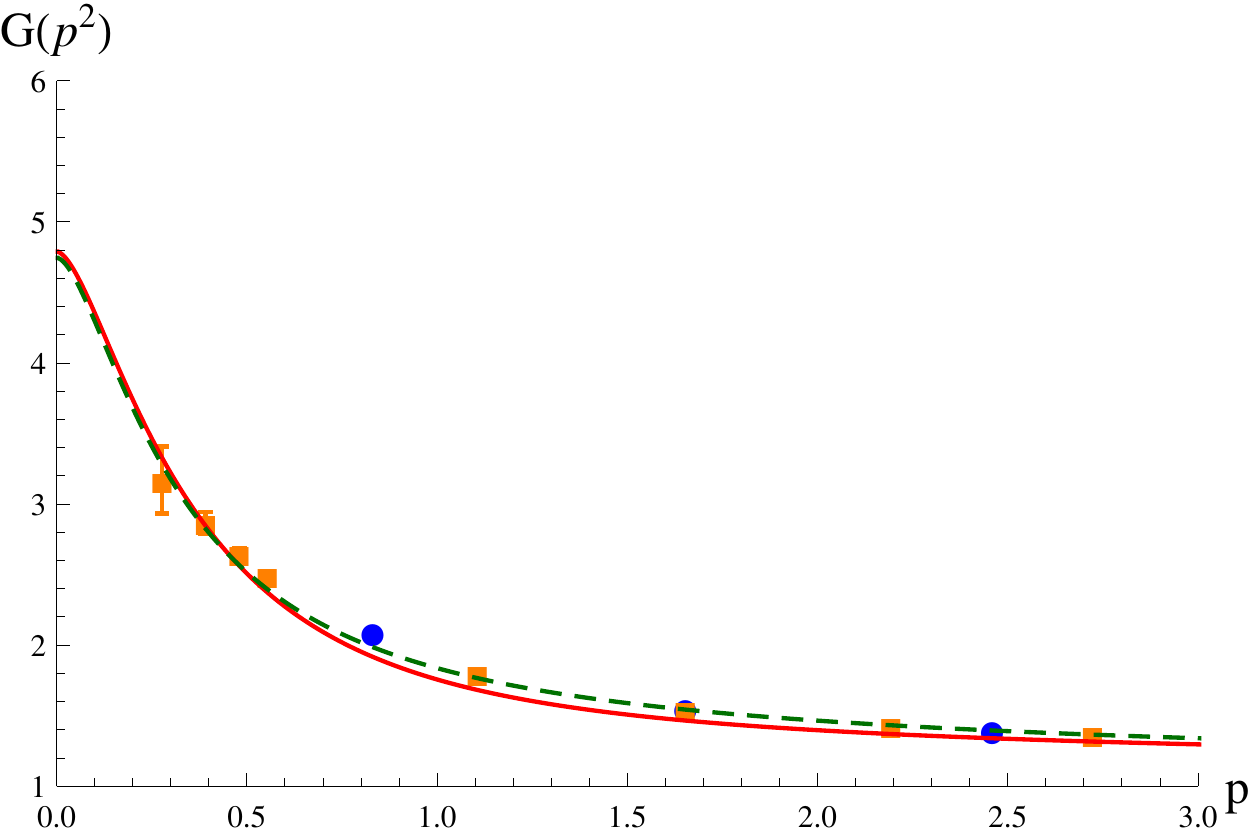}
  \caption{\label{fig:compLDynGhg_props}Comparison of the results for the gluon propagator dressing function $Z(p^2)$ and the ghost dressing function $G(p^2)$ with lattice data \cite{Sternbeck:2006rd}. The red/continuous line was obtained with a dynamic ghost-gluon vertex and the optimized effective three-gluon vertex with $h_{IR}=-1$ and $\Lambda_{3g}=2.1\,GeV$. For reference the green/dashed line is shown, which was obtained with the vertex models of refs.~\cite{Fischer:2002hn,Fischer:2002eq} and a bare ghost-gluon vertex. Lattice data \cite{Sternbeck:2006rd} is for $\beta=6$ and lattice sizes of $L=32$ (blue circles) and $L=48$ (orange squares).
}
 \end{center}
\end{figure}

\begin{figure}
 \begin{center}
  \includegraphics[width=0.49\textwidth]{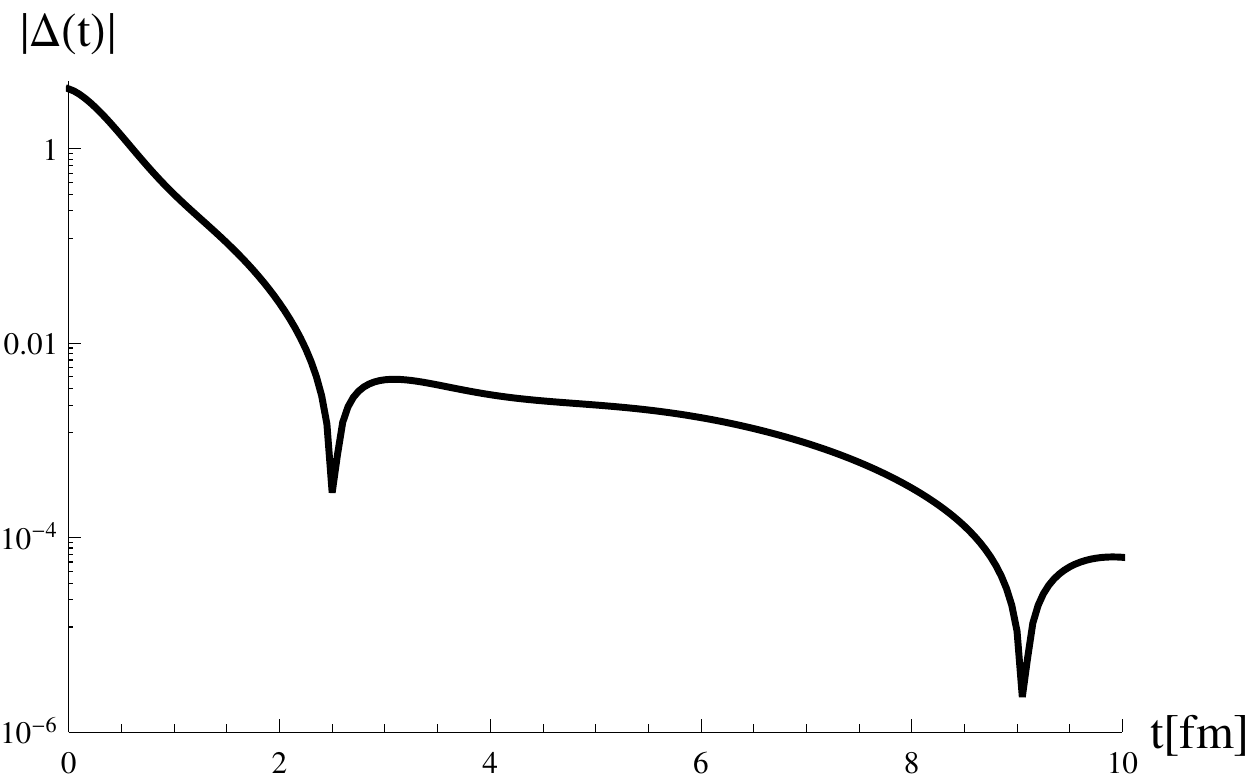}
  \includegraphics[width=0.49\textwidth]{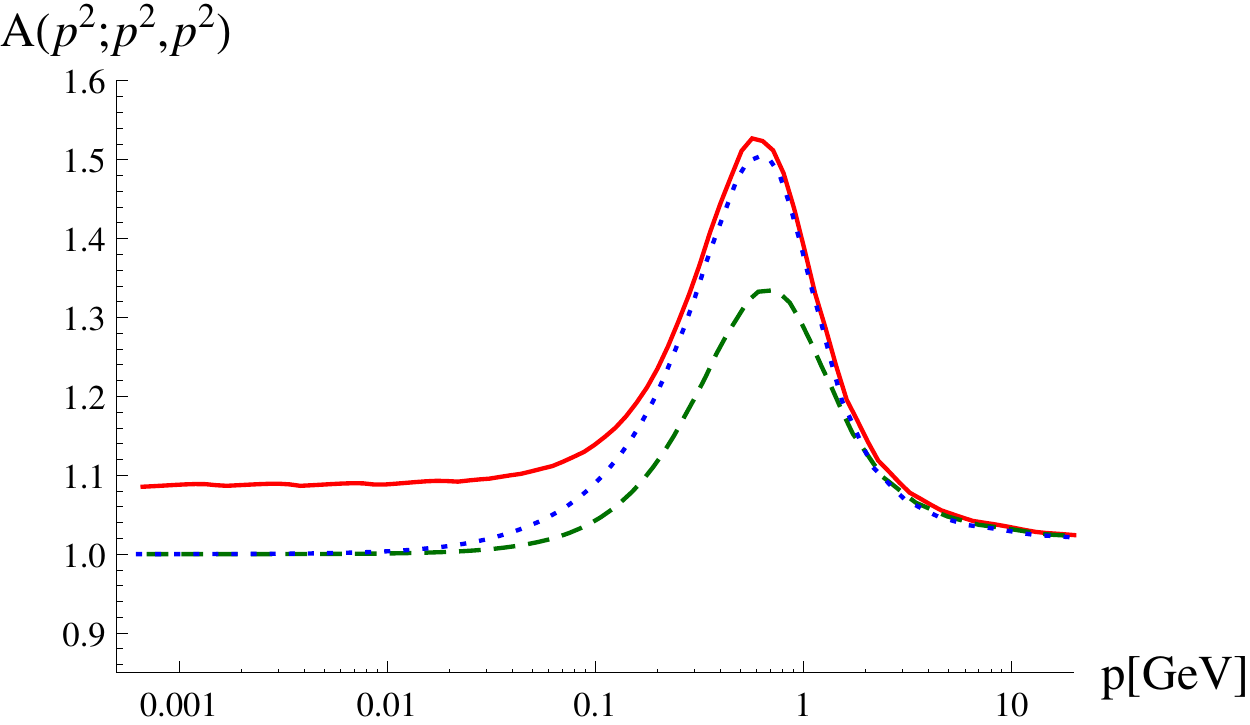}
 \end{center}
 \caption{\label{fig:SchwingerFunc-GhgDecScal}\textit{Left}: The absolute value of the Schwinger function of the gluon propagator. In the region between approximately $2.2\,GeV$ and $9\,GeV$ the function is negative. \textit{Right}: The ghost-gluon vertex at the symmetric point. The red/continuous line is from the scaling solution, and the green/dashed and blue/dotted lines from two different decoupling solutions.}
\end{figure}

The resulting dressing function for the propagators are shown in \fref{fig:compLDynGhg_props}. As can be seen, it is possible to choose the parameter $\Lambda_{3g}$ such that we obtain a very good agreement with lattice data. We would like to stress that this is true for \emph{both} dressings. The Schwinger function $\Delta(t)$ of the gluon propagator is shown in \fref{fig:SchwingerFunc-GhgDecScal}. It turns negative at about $2.2\,fm$. This value is higher than observed in previous calculations \cite{Fischer:2008uz}. Furthermore, a second zero crossing is observed. Since the Schwinger function is partially negative, positivity is violated and the gluon cannot represent an asymptotic state of the physical state space.

\begin{figure}[tb]
 \begin{center}
 \begin{minipage}[t]{0.48\textwidth}
  \includegraphics[width=\textwidth]{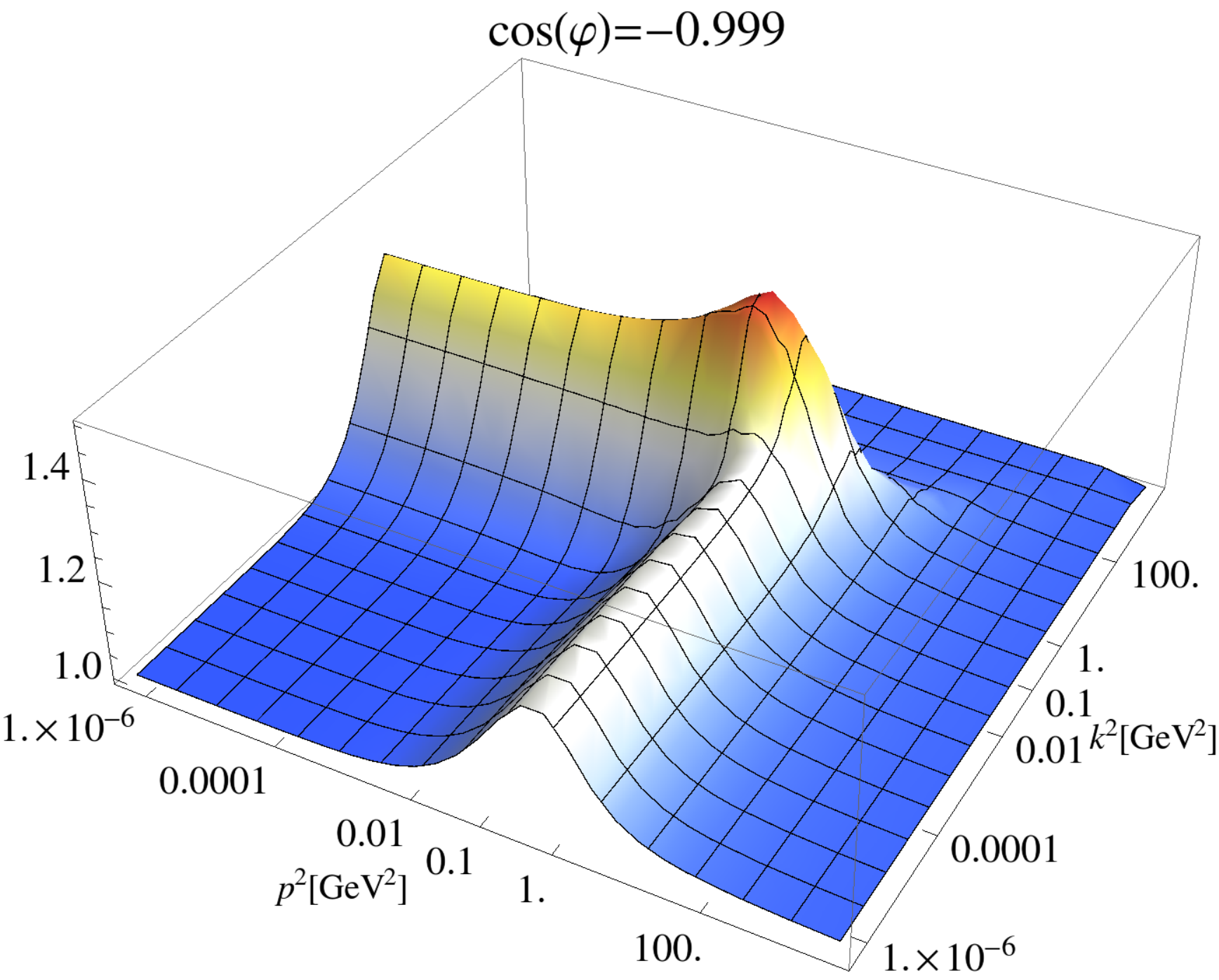}
 \end{minipage}
 \hfill
 \begin{minipage}[t]{0.48\textwidth}
  \includegraphics[width=\textwidth]{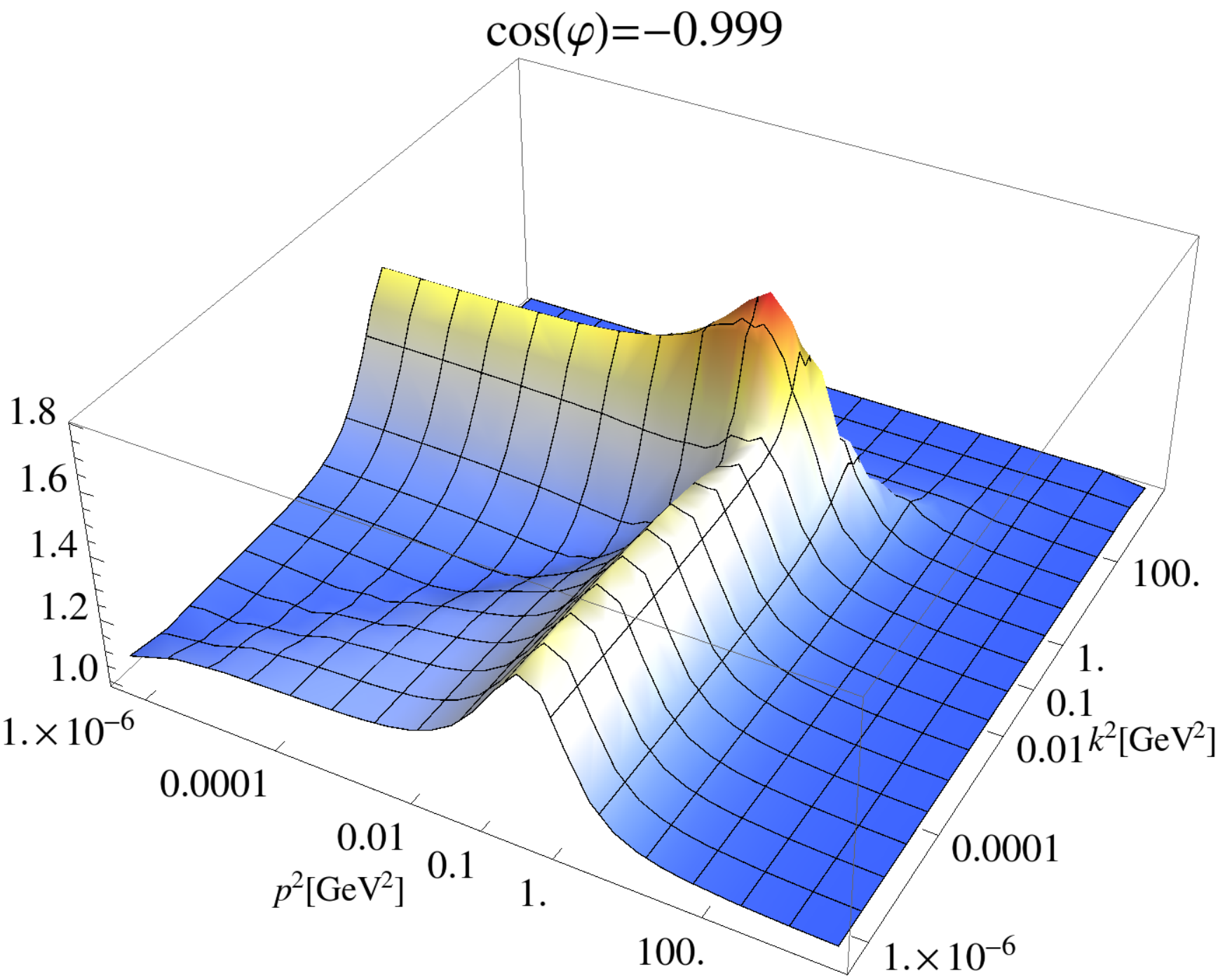} 
 \end{minipage}
 \caption{\label{fig:dynGhg_ghg}Ghost-gluon vertex dressing with almost anti-parallel gluon and anti-ghost momenta for decoupling (\textit{left}) and scaling (\textit{right}).}
 \end{center}
\end{figure}

Results for the ghost-gluon vertex are shown in \fref{fig:dynGhg_ghg}. The depicted momentum configuration corresponds to almost anti-parallel gluon and anti-ghost momenta. As expected the dressing is constant in the IR and the UV and has a small bump in the intermediate momentum regime. The ambiguity between scaling and decoupling is also present in the vertex, as the IR constant is different: For decoupling the dressing is $1$, because the loop corrections are IR suppressed by the gluon propagators. For scaling, however, they yield a non-vanishing contribution which lifts the dressing above $1$. This can also be seen from the red/continuous line in \fref{fig:SchwingerFunc-GhgDecScal}, where additionally two different decoupling solutions are plotted.

For the scaling solution the exponents of the IR power laws for all dressing functions can be calculated in terms of one variable $\kappa$ \cite{Alkofer:2004it,Fischer:2009tn}, which can be calculated analytically \cite{Zwanziger:2001kw,Lerche:2002ep}. Under the assumption that the IR limit of the ghost-gluon vertex is regular, its value is $\kappa=0.595353$. This value can only be modified by a non-analyticity at zero momentum \cite{Lerche:2002ep}. One possibility for this, a dependence on the angle between two momenta, was studied in ref.~\cite{Huber:2012zj}. Here we tried to find a deviation by not fixing the value of $\kappa$ manually, as is normally done, but leaving it free. Thus the ghost-gluon vertex could develop an angle dependence, which was, however, not observed. The differences between the vertices as obtained from both procedures were below $0.1\%$ and thus within numerical accuracy. The value of $\kappa$ was found to be a little bit higher, slightly above $0.6$. Thus we did not find any non-analyticities in the ghost-gluon vertex.

\section{Summary and conclusions}

In general DSEs are a useful tool for non-perturbative investigations. For Landau gauge Yang-Mills theory we shortly recalled the history of different truncations and explicitly demonstrated that further improving the truncation leads only to quantitative effects. One improvement presented here is the dynamical inclusion of the ghost-gluon vertex. We find that the ambiguity of decoupling and scaling solutions persists also at the level of three-point functions. We also introduced a model for the three-gluon vertex that properly describes lattice data. This model may also be relevant for future calculations when the so-called squint diagram is included into the calculation. Results for the sunset were presented in \cite{Mader:2013cx}. In summary, we could reproduce both propagators as obtained from Monte Carlo simulations within the present setup.

\section*{Acknowledgments}
M.Q.H.\ was supported by the Alexander von Humboldt foundation, L.v.S.\ by the Helmholtz International Center for FAIR within the LOEWE program of the State of Hesse, the Helmholtz Association Grant VH-NG-332, and the European Commission, FP7-PEOPLE-2009-RG No.~249203. Images were created with \textit{Mathematica} \cite{Wolfram:2004}.

\bibliographystyle{utphys_mod}
\bibliography{literature_YM4d_ConfinementX}

\end{document}